\documentclass[preprint,nofootinbib]{revtex4}
\usepackage{amssymb}
\usepackage{graphics}

\usepackage{multirow}
\usepackage{array} 

\usepackage{hyperref}

\usepackage{amssymb}
\usepackage{times,fancyhdr}
\usepackage{amsmath}
\usepackage{amsfonts}
\usepackage{epstopdf}

\begin{document}

%............................ definitions ..............
\newcommand{\be}{\begin{equation}}
\newcommand{\ee}{\end{equation}}
\def\bq{\begin{eqnarray}}
\def\eq{\end{eqnarray}}
%........................................................

\title{\bf A New Weyl-like Tensor of Geometric Origin}
\author{Ram Gopal Vishwakarma\footnote{Email: vishwa@uaz.edu.mx}}

 \address{Unidad Acad$\acute{e}$mica de Matem$\acute{a}$ticas\\
 Universidad Aut$\acute{o}$noma de Zacatecas\\
 C.P. 98068, Zacatecas, ZAC, Mexico}

\begin{abstract}
A set of new tensors of purely geometric origin have been investigated, which form a hierarchy.  A tensor of a lower rank plays the role of the potential for the tensor of one rank higher. The tensors have interesting mathematical and physical properties. The highest rank-tensor of the hierarchy, possesses all the geometrical properties of the Weyl tensor.
\end{abstract}



\maketitle

\section{Introduction}

Riemannian geometry, originated from the deep insight and revolutionary vision of Georg Bernhard Riemann,  has two salient features - curvature and symmetries - which can exist, in general, independently of each other. 
The former facilitated  the formulation of general relativity (GR) - Einstein's theory of gravitation - in terms of the curvature of the four-dimensional spacetime.
The latter - the symmetries in spacetime -  have been extensively studied from both, the mathematical and the physical viewpoints. 

The Riemann-Christoffel curvature tensor is the nodal point for the unfolding of GR. However, there is another tensor that in some ways is more
fundamental. It is the Weyl conformal tensor.
The central objective of this paper is to develop and study a new set of tensors emerging purely from the symmetries of the spacetime. Interestingly, one of the tensors seems to possess all the algebraic properties of the Weyl tensor.

\section{A New Set of Tensors}

The geometry associated with a Riemannian manifold is usually developed by considering vectors as the fundamental geometrical objects. 
Let us consider a spacetime admitting a Killing vector field $K^\alpha$. As a Killing vector field preserves the metric along its direction, it satisfies the Killing equation
\be
K_{\mu;\nu}+K_{\nu;\mu}=0,\label{eq:Killing}
\ee
where the semicolon followed by an index denotes covariant derivative (associated with the metric $g_{\mu\nu}$) with respect to the corresponding variable. (Sometimes we shall denote the covariant derivative with $\nabla$ also.)
Let us define 
\be
F_{\mu\nu}\equiv \frac{1}{2}(K_{\mu;\nu}-K_{\nu;\mu}),\label{eq:F}
\ee
\be
U_{\alpha\kappa\mu}\equiv F_{\alpha\kappa;\mu}, ~~ M_{\alpha\kappa\mu\nu}\equiv U_{\alpha\kappa\mu;\nu},  ~~ N_{\alpha\kappa\mu\nu}\equiv  M_{[\alpha\kappa][\mu\nu]} + M_{[\mu\nu][\alpha\kappa]},  \label{eq:U}
\ee
\be
V_{\alpha\kappa\mu\nu}\equiv N_{\alpha\kappa\mu\nu}-{*N*}_{\alpha\kappa\mu\nu}= U_{[\alpha\kappa][\mu;\nu]} + U_{[\mu\nu][\alpha;\kappa]} -  {*U*}_{[\alpha\kappa][\mu;\nu]} - {*U*}_{[\mu\nu][\alpha;\kappa]},\label{eq:V}
\ee
where the square brackets [] enclosing indices denote the antisymmetrized sum over the bracketed indices and the starred symbol denotes its dual defined by ${*N*}_{\alpha\kappa\mu\nu}=\frac{1}{4}e_{\alpha\kappa\rho\sigma}e_{\mu\nu\tau\delta} N^{\rho\sigma\tau\delta}$, with $e_{\alpha\kappa\gamma\delta}$ representing the Levi-Civita tensor (see Appendix A).

As the tensors $F_{\mu\nu}$ and $V_{\mu\nu\alpha\kappa}$ are defined in terms of the gradients of respectively $K_\alpha$ and $U_{\mu\nu\alpha}$, the latter two respectively appear as the potentials for the formers. For the same reason, $F_{\mu\nu}$ appears as the potential for $U_{\mu\nu\alpha}$. The definition (\ref{eq:F}) implies that $F_{\mu\nu}$ admits the following symmetries
\be
F_{\mu\nu}=-F_{\nu\mu}, ~~~~~ F_{\mu\nu;\alpha}+F_{\nu\alpha;\mu}+F_{\alpha\mu;\nu}=0,\label{eq:I-F}
\ee
by virtue of which, the definition (\ref{eq:U}) supplies the following two symmetries for the tensor $U_{\mu\nu\alpha}$
\be
U_{\mu\nu\alpha}= - U_{\nu\mu\alpha}, ~~~~~ U_{\mu\nu\alpha}+U_{\nu\alpha\mu}+U_{\alpha\mu\nu}=0.\label{eq:I-U}
\ee
From the definition (\ref{eq:V}), it is readily apparent that the tensor $V_{\mu\nu\alpha\kappa}$ has the following three symmetries
\be
V_{\mu\nu\alpha\kappa}=-V_{\nu\mu\alpha\kappa}=-V_{\mu\nu\kappa\alpha}=V_{\alpha\kappa\mu\nu},\label{eq:I-V}
\ee 
which are also shared by $N_{\mu\nu\alpha\kappa}$. By using the Lanczos identity (see Appendix A), the duals appearing in (\ref{eq:V}) can easily be evaluated and (\ref{eq:V}) can be written in the form
\begin{eqnarray}\nonumber
V_{\alpha\kappa\mu\nu} &=& U_{\alpha\kappa\mu;\nu}-U_{\alpha\kappa\nu;\mu}+U_{\mu\nu\alpha;\kappa}-U_{\mu\nu\kappa;\alpha}
+ g_{\kappa\mu}U_{(\alpha\nu)}+g_{\alpha\nu}U_{(\kappa\mu)}\\
&&-g_{\kappa\nu}U_{(\alpha\mu)}-g_{\alpha\mu}U_{(\kappa\nu)}
+\frac{2}{3}U^{\rho\sigma}_{~~~\rho;\sigma}\{g_{\alpha\mu}g_{\kappa\nu} - g_{\alpha\nu}g_{\kappa\mu}\}, \label{eq:VV}
\end{eqnarray}
where $U_{\alpha\nu}\equiv U_{\alpha~\nu;\sigma}^{~\sigma}-U_{\alpha~\sigma;\nu}^{~\sigma}$ and the round brackets () enclosing indices, denote the symmetrized sum over the bracketed indices.
By writing the cyclic sum $V_{\alpha\kappa\mu\nu}+V_{\alpha\mu\nu\kappa}+V_{\alpha\nu\kappa\mu}$ from equation (\ref{eq:VV}) and using therein the cyclic identity for $U_{\alpha\kappa\mu}$ shown in (\ref{eq:I-U}), we find that $V_{\mu\nu\alpha\kappa}$ also admits the cyclic identity
\be
V_{\alpha\kappa\mu\nu}+V_{\alpha\mu\nu\kappa}+V_{\alpha\nu\kappa\mu}=0.\label{eq:CI-V}
\ee
Similarly, one can check from (\ref{eq:VV}) that the tensor $V_{\mu\nu\alpha\kappa}$ is traceless in all pairs of indices and satisfies
\be
{*V}_{\mu\nu\alpha\kappa}={V*}_{\mu\nu\alpha\kappa}. \label{eq:LRV1}
\ee
The symmetries of this tensor (and its dual), discovered in (\ref{eq:I-V}), (\ref{eq:CI-V}), (\ref{eq:LRV1}), (\ref{eq:Vd1})-(\ref{eq:VCyclic}),  are summarized in Table \ref{table:sym}, which asserts that the new tensor $V_{\mu\nu\alpha\kappa}$ possesses all the algebraic properties of the Weyl tensor.

\begin{table}
\caption{\bf Algebraic symmetries of $V_{\mu\nu\alpha\kappa}$ and its duals} 
\centering
\begin{tabular}{|  c c c  c |} 
\hline\hline
 Symmetry ~~ ~ & Antisymmetry  ~~~ & Cyclicity  ~~~ & Trace-freeness \\ [0.5ex] % [0.5ex] adds vertical space
\hline 

 $V_{\mu\nu\alpha\kappa}=V_{\alpha\kappa\mu\nu}$ ~~~~ &  $V_{\mu\nu\alpha\kappa}=V_{[\mu\nu][\alpha\kappa]}$ ~~~~  &  $V_{\mu[\nu\alpha\kappa]}=0$ ~~~~   &   $V^\nu_{~~\mu\nu\alpha}=0$ \\  [2ex]

&   ~~~~~~~ ${*V}_{\mu\nu\alpha\kappa}={V*}_{\mu\nu\alpha\kappa}\equiv \overset{*}{V}_{\mu\nu\alpha\kappa} \Rightarrow$ & &\\ [1ex]
 
~$\overset{*}{V}_{\mu\nu\alpha\kappa}=\overset{*}{V}_{\alpha\kappa\mu\nu}$  ~~~  &   $\overset{*}{V}_{\mu\nu\alpha\kappa}=\overset{*}{V}_{[\mu\nu][\alpha\kappa]}$ ~~~~  &  $\overset{*}{V}_{\mu[\nu\alpha\kappa]}=0$ ~~~~   &   $\overset{*}{V^\nu}_{\mu\nu\alpha}=0$ \\ [1ex] 
\hline 
\end{tabular}
\label{table:sym} % is used to refer this table in the text
\end{table}

\section{Physical Interpretation of the New Tensors}

So far we have not utilized the primary assumption (\ref{eq:Killing}) of $K^\mu$  being a Killing vector field. Hence everything we have discussed so far is valid for $K^\mu$ as an arbitrary vector field. However, restricting $K^\mu$  to Killing fields, garnishes with interesting physical properties to some of the newly-defined tensors, as we shall see in the following.

\subsection{Recovering Maxwell Equations}

By the use of (\ref{eq:Killing}), the tensors $F_{\mu\nu}$ and  $U_{\mu\nu\kappa}$ reduce to the simple form 
\be
 F_{\mu\nu} = K_{\mu;\nu}, ~~~~ U_{\mu\nu\kappa}= K_{\mu;\nu;\kappa}. \label{eq:FK}
\ee
The tensor $K^\mu$  can also be linked with the Riemann tensor through its defining equation 
\[
K_{\alpha;\nu;\mu}-K_{\alpha;\mu;\nu}= R_{\mu\nu\alpha\sigma} K^\sigma,\label{eq:Riemann}
\]
which, by the use of  (\ref{eq:Killing}), reduces to 
\[
K_{\alpha;\nu;\mu} + K_{\mu;\alpha;\nu} = R_{\mu\nu\alpha\sigma} K^\sigma.
\]
Following \cite{Wald}, let us add to this equation the one obtained by the swap of indices $(\mu, \nu,\alpha)\rightarrow(\nu, \alpha, \mu)$  and then subtrac the one obtained from the swap of indices $(\nu, \alpha, \mu)\rightarrow(\alpha, \mu, \nu )$. By using the cyclic symmetry of the Riemann tensor in the obtained result,  we get 
\[
K_{\mu;\alpha;\nu} + R_{\alpha\mu\nu\sigma}K^\sigma =0.
\]
By contracting over $\nu$ and $\alpha$, this yields a wave equation for $K^\mu$:
\be
\nabla^\nu \nabla_\nu K_\mu \equiv  g^{\alpha\nu} K_{\mu;\alpha;\nu} =- R_{\mu\sigma}K^\sigma.\label{eq:WaveK}
\ee
In the Ricci-flat case ($R_{\mu\nu}=0$), this reduces to the homogeneous wave equation
\be
\nabla^\nu \nabla_\nu K_\mu =0.\label{eq:Max}
\ee
As is shown in the following, this is, in disguise, the source-free Maxwell equations in Lorenz gauge (the Lorenz gauge condition  $K^\sigma_{~;\sigma}=0$ follows immediately from (\ref{eq:Killing}) by contracting over $\mu$ and $\nu$). Equation (\ref{eq:FK}) implies that
\be
\nabla^\nu F_{\mu\nu}=\nabla^\nu \nabla_\nu K_\mu=0,\label{eq:Max2}
\ee
by virtue of  (\ref{eq:Max}). Let us note that equations  (\ref{eq:I-F}) and (\ref{eq:Max2}) constitute the source-free Maxwell equations. Thus the tensors $F_{\mu\nu}$ and $K^\mu$ become respectively the electromagnetic field tensor and its potential  in a Ricci-flat spacetime. 
It has already been noted that any Killing vector field present in the Ricci-flat spacetime, gives rise to a solution of Maxwell equations \cite{Wald}. In this case, the tensor $U_{\mu\nu\kappa}$ admits an additional symmetry
\be
U_{\mu~\nu}^{~\nu}=0, \label{eq:USym}
\ee
by virtue of (\ref{eq:Max2}).

\subsection{Gauge-Invariance }

One can check that the tensor $F_{\mu\nu}$ defined by (\ref{eq:F}) remains unchanged under a gauge transformation 
\be
\bar{K}_\mu=K_\mu+\phi_\mu.
\ee
That is,  $F_{\mu\nu}$ does not change if we add an arbitrary gradient $\phi_\mu\equiv\phi_{; \mu}$ to the potential $K_\mu$, where $\phi$ is an arbitrary scalar function.  This happens because
\[
\bar{F}_{\mu\nu}=\frac{1}{2}[(K_{\mu}+\phi_\mu)_{;\nu}-(K_{\nu}+\phi_\nu)_{;\mu}]=F_{\mu\nu}
\]
by noting that the Christoffel symbols cancel out and the partial derivatives commute \cite{ABS}. 
In a similar fashion, the tensor $V_{\mu\nu\kappa\alpha}$ is invariant under the gauge transformation 
\be
\bar{U}_{\mu\nu\kappa} = U_{\mu\nu\kappa}+ X_\mu g_{\nu\kappa}- X_\nu g_{\mu\kappa}, \label{eq:gaugeU}
\ee
where $ X_\alpha$ is an arbitrary vector field. This is readily verified by substituting (\ref{eq:gaugeU}) in (\ref{eq:VV}). A straightforward calculation then shows that $\bar{V}_{\mu\nu\kappa\alpha}=V_{\mu\nu\kappa\alpha}$. [This reminds us of the Weyl tensor which is invariant under the transformation (\ref{eq:gaugeU}) performed on its potential tensor - the Lanczos tensor \cite{Novello}. We shall return to this issue in section  \ref{sec:NWeyl}.] Let us note that this gauge-invariance of the tensors $F_{\mu\nu}$ and $V_{\mu\nu\kappa\alpha}$ holds even when $R_{\mu\nu}\neq 0$.

In the Ricci-flat case, where the tensors $F_{\mu\nu}$ and  $K_\mu$ play the roles of the electromagnetic field tensor and its potential respectively, this arbitrariness in $K_\mu$ is removed with the aid of the Lorenz gauge condition (constraining $\phi$ with $\nabla^\nu \nabla_\nu \phi =0$). The 
arbitrariness in $U_{\mu\nu\kappa}$ is automatically removed in the  Ricci-flat case with the aid of symmetry (\ref{eq:USym}), which implies a vanishing $X_\mu$.

\subsection{Scale-Invariance }

It is well-known that  the Weyl tensor and the source-free Maxwell equations are invariant under a scale change of the metric \cite{ABS}:
\be
\bar{g}_{\mu\nu}=A(x^\alpha) g_{\mu\nu},\label{eq:gauge}
\ee 
where $A(x^\alpha)$ is an arbitrary function of $x^\alpha$. Although we could not verify this invariance for $V_{\mu\nu\kappa\alpha}$ for an arbitrary $A(x^\alpha)$ (where the calculations become cumbersome), however it is easy to show that this holds for a constant $A$. 

Nevertheless, we  predict that a similar relation, like that for Weyl, must exist for $V_{\mu\nu\kappa\alpha}$ for the general case (\ref{eq:gauge}). Our assertion is based on an explicit calculation of the tensor $V_{\mu\nu\kappa\alpha}$ in some conformally flat spacetimes - Robertson-Walker, interior-Schwarzschild and de Sitter - by using the definitions (\ref{eq:Killing})-(\ref{eq:V}). The tensor vanishes identically! (Let us note that the tensor $F_{\mu\nu}$ becomes constant and hence the tensors $U_{\mu\nu\alpha}$ and $V_{\mu\nu\kappa\alpha}$ vanish for the flat maximally symmetric Minkowskian spacetime with ten Killing vectors.)

\section{A Bel-Robinson-like Tensor from $V_{\mu\nu\kappa\alpha}$}

Continuing our exploration of the similarities between the the Weyl tensor and  $V_{\mu\nu\kappa\alpha}$, we next find that the latter also supports a completely symmetric and trace-free rank-4 tensor, which vanishes if and only if the tensor $V_{\mu\nu\kappa\alpha}$ vanishes.

Exploring the analogy between the electromagnetic and the gravitational fields, Bel \cite{B} discovered a rank-4 tensor, constructed from the Weyl
tensor, which is analogous to the energy-momentum tensor of the electromagnetic field. The tensor, often called the Bel-Robinson tensor, is defined
for every spacetime (even for non-Ricci-flat ones \cite{B-R}) as
\be
 T_{\alpha\beta\gamma\delta} = C^{~~\sigma\rho}_{\alpha~~~\gamma}~ C_{\beta\sigma\rho\delta} + *C^{~~\sigma\rho}_{\alpha~~~\gamma}~ {*C}_{\beta\sigma\rho\delta},\label{eq:B-R}
\ee
where $C_{\mu\nu\kappa\alpha}$ is the conformal Weyl tensor. As the tensor $V_{\mu\nu\kappa\alpha}$ shares all the algebraic symmetries of the Weyl tensor, it can also support a tensor formulated along the lines of (\ref{eq:B-R}):
\be
 W_{\mu\nu\kappa\alpha} = V^{~\sigma\rho}_{\mu~~~\kappa}~ V_{\nu\sigma\rho\alpha} + *V^{~\sigma\rho}_{\mu~~~\kappa}~ {*V}_{\nu\sigma\rho\alpha}, \label{eq:W}
\ee
which, by the use of (\ref{eq:C1}), can alternatively be written as 
\be
W_{\mu\nu\kappa\alpha} = V_{\mu~~~\kappa}^{~\sigma\rho} ~ V_{\nu\sigma\rho\alpha} + V_{\mu~~~\alpha}^{~\sigma\rho} ~ V_{\nu\sigma\rho\kappa}  - \frac{1}{2}g_{\mu\nu}~V_{\kappa\sigma\rho\tau}V_{\alpha}^{~\sigma\rho\tau}. \label{eq:WW}
\ee
By using the symmetry properties of $V_{\mu\nu\kappa\alpha}$ in (\ref{eq:W}) and (\ref{eq:WW}) and using  (\ref{eq:C6}), one can check that
\[
W_{\mu\nu\kappa\alpha} =W_{(\mu\nu)(\kappa\alpha)} =W_{\kappa\alpha\mu\nu}=W_{\kappa\nu\mu\alpha}.
\]
By virtue of these symmetries, one arrives at the conclusion that $W_{\mu\nu\kappa\alpha}$ is symmetric in all pairs of indices. This, taken together with the symmetries of $V_{\mu\nu\kappa\alpha}$, implies that the contraction of any two indices of $W_{\mu\nu\kappa\alpha}$ vanishes identically. Thus the tensor $W_{\mu\nu\kappa\alpha}$ possesses all the symmetries of the tensor $T_{\alpha\beta\gamma\delta}$ defined in  (\ref{eq:B-R}).
Interestingly, if we formulate, along the lines of (\ref{eq:W}), a rank-2 tensor from $V_{\mu\nu\kappa\alpha}$, i.e., 
\be
 W_{\mu\nu} = V_\mu^{~\sigma\rho\kappa}~ V_{\nu\sigma\rho\kappa} + {*V}_\mu ^{~~\sigma\rho\kappa} ~ {*V}_{\nu\sigma\rho\kappa}, \label{eq:W-2}
\ee
the new tensor vanishes identically, as $ {*V}_\mu ^{~~\sigma\rho\kappa} ~ {*V}_{\nu\sigma\rho\kappa}=-V_\mu^{~\sigma\rho\kappa}~ V_{\nu\sigma\rho\kappa}$ (which can be arrived at by replacing the left duals by right duals and then following the steps of  (\ref{eq:C5}) to evaluate them).

\section{$V_{\mu\nu\kappa\alpha}$ is Distinct from the Weyl Tensor} 
\label{sec:NWeyl}

The tensor $V_{\mu\nu\kappa\alpha}$ may be taken for the Weyl tensor, due to its striking similarities with the latter and its geometric origin,
particularly when it originates from the Killing vector fields and thus gets strongly coupled with the geometry of the spacetime.  In this case,  the potential tensor $U_{\mu\nu\kappa}$ would be taken for the Lanczos tensor. Nevertheless, despite their geometric origin and symmetries, the tensors $U_{\mu\nu\kappa}$ and $V_{\mu\nu\kappa\alpha}$ appear distinct from the Lanczos and  the Weyl tensors in general, as we shall see in the following. Before coming to this point however, it would be worthwhile to take a digression and recapitulate the Lanczos tensor, as this subject could not gain proper attention in the mainstream GR research and has remained more or less a mathematical curiosity only.

The existence of a rank-3 tensor $L_{\mu\nu\alpha}$ serving as the potential for the Weyl tensor in 4-dimensional spacetimes was first proposed by Lanczos \cite{Lanczos}. Lanczos showed that the tensor $L_{\mu\nu\alpha}$ satisfying
\be
L_{\mu\nu\alpha}+ L_{\nu\mu\alpha}=0,        ~~~~~~~ \text{(a)}~~~~~    L_{[\mu\nu\alpha]}=0 ~~~(\text{or equivalently}~ {*L}_{\mu ~\nu}^{~\nu}=0), ~~~~~~ \text{(b)}  \label{eq:S-L}
\ee
could generate the Weyl tensor $C_{\mu\nu\kappa\alpha}$ through equation
\be
C_{\alpha\kappa\mu\nu} = L_{[\alpha\kappa][\mu;\nu]} + L_{[\mu\nu][\alpha;\kappa]} -  {*L*}_{[\alpha\kappa][\mu;\nu]} - {*L*}_{[\mu\nu][\alpha;\kappa]},\label{eq:C-L}
\ee
and hence through
\begin{eqnarray}\nonumber
C_{\alpha\kappa\mu\nu} &=& L_{\alpha\kappa\mu;\nu}-L_{\alpha\kappa\nu;\mu}+L_{\mu\nu\alpha;\kappa}-L_{\mu\nu\kappa;\alpha}
+ g_{\kappa\mu}L_{(\alpha\nu)}+g_{\alpha\nu}L_{(\kappa\mu)} - g_{\kappa\nu}L_{(\alpha\mu)}\\
&&-g_{\alpha\mu}L_{(\kappa\nu)}+\frac{2}{3}L^{\rho\sigma}_{~~\rho;\sigma}\{g_{\alpha\mu}g_{\kappa\nu} - g_{\alpha\nu}g_{\kappa\mu}\},~~\text{with} ~~ L_{\alpha\nu}\equiv L_{\alpha~\nu;\sigma}^{~\sigma}-L_{\alpha~\sigma;\nu}^{~\sigma}. 
 \label{eq:CC-L}
\end{eqnarray}
(Let us note that these equations are nothing but equations (\ref{eq:V}) and (\ref{eq:VV}) with $U$ and $V$ replaced by $L$ and $C$ respectively.)
The method employed by Lanczos is the variational process wherein (\ref{eq:C-L}) appears as the Euler-Lagrange equation of a suitable functional. Thenceforth, rigorous proofs of the existence of the potential $L_{\mu\nu\alpha}$ for the Weyl tensor have been provided in \cite{B&C,Illge}.
It has also been shown that this potential tensor does not generally exist for dimensions higher than four \cite{E-H2}. Interestingly, the potential tensor exist only for the Weyl tensor, and not for the Riemann tensor in general \cite{Edgar}.

By noticing that the Weyl tensor has only 10 degrees of freedom, whereas the tensor $L_{\mu\nu\alpha}$ obeying constraints (\ref{eq:S-L}) has 20 independent components, Lanczos has imposed two additional constraints
\be
L_{\mu ~~\nu}^{~~\nu}=0,       ~~~~~~~ ~~~~ \text{(a)}     ~~~~~ ~~~  L_{\mu\nu~;\sigma}^{~~~\sigma}=0, ~~~~~~~~~~ \text{(b)}  \label{eq:SS-L}
\ee
supplying 10 equations to eliminate the excess degrees of freedom. While Lanczos appears to use conditions (\ref{eq:SS-L}) to reduce the excess degrees of freedom, many authors appear to use them as a part of the definition of $L_{\mu\nu\alpha}$, which creates a confusing picture especially for the beginners  as to what the minimum requirements are for the existence of the potential tensor $L_{\mu\nu\alpha}$.
This issue has also been addressed in \cite{E-H1}.

The reference \cite{B&C} however provides not only a valid and rigorous proof of existence of $L_{\mu\nu\alpha}$ but also shows that the conditions  (\ref{eq:SS-L}) are not essential for the existence of $L_{\mu\nu\alpha}$, and  any values on the right hand sides of the two
equations in (\ref{eq:SS-L}) are admissible.
Therefore these conditions  remain as arbitrary gauge choices - the Lanczos algebraic gauge and the Lanczos differential gauge.

There is yet another issue related with the conventional portrayal of $L_{\mu\nu\alpha}$ in the literature, which creates confusion about the role of condition (\ref{eq:S-L}.b) in equation (\ref{eq:C-L}). The confusion arises from the claim that ${*C}_{\mu\nu\alpha\kappa}={C*}_{\mu\nu\alpha\kappa}$ is a consequence of the properties 
\be
C_{\mu\nu\alpha\kappa}=C_{[\mu\nu][\alpha\kappa]}=C_{[\alpha\kappa][\mu\nu]}, ~ C^\alpha_{~\mu\alpha\nu}=0,   \label{eq:CS1}
\ee
taken together with 
\be
C_{\mu[\nu\alpha\kappa]}=0,  \label{eq:CS2}
\ee 
(which is though not correct. We notice, in Appendix B, that the last symmetry ``$C_{\mu[\nu\alpha\kappa]}=0$'' is {\it not} required to have ${*C}_{\mu\nu\alpha\kappa}={C*}_{\mu\nu\alpha\kappa}$. It is rather ${*C}_{\mu[\nu\alpha\kappa]}=0$, which is instrumental to have ${*C}_{\mu\nu\alpha\kappa}={C*}_{\mu\nu\alpha\kappa}$.)
As  equations (\ref{eq:C-L}), (\ref{eq:CC-L}) readily hold ${*C}_{\mu\nu\alpha\kappa}={C*}_{\mu\nu\alpha\kappa}$, the mentioned ``claim'' makes one expect the symmetry (\ref{eq:CS2}) to appear from (\ref{eq:CC-L}), in addition to the symmetries (\ref{eq:CS1}), {\it without requiring any other condition but} (\ref{eq:S-L}.a).  Let us note that it is only the condition (\ref{eq:S-L}.a), {\it and not} (\ref{eq:S-L}.b), which is required to derive (\ref{eq:CC-L}) from (\ref{eq:C-L}). What would then be left for the condition (\ref{eq:S-L}.b) to have any role in the definition of  $L_{\mu\nu\alpha}$? Though the symmetries (\ref{eq:CS1}) are indeed admitted by (\ref{eq:CC-L}) (when it is taken together with  (\ref{eq:S-L}.a)), but not the symmetry (\ref{eq:CS2}) unless one uses (\ref{eq:S-L}.b)) therein, as can be checked in a direct calculation.  So, here is the role of the condition (\ref{eq:S-L}.b) that upholds  (\ref{eq:CS2})!

\subsection{Example 1}

Having given this brief overview of the Lanczos tensor, it would now be easy to show that the tensors $U_{\mu\nu\kappa}$ and $V_{\mu\nu\kappa\alpha}$ are distinct from the Lanczos and Weyl tensors. This can be shown by considering some examples. Let us first consider, for instance, the Schwarzschild solution
\be
ds^2=\left(1-\frac{2m}{r}\right) dt^2-\frac{dr^2}{(1-2m/r)}-r^2d\theta^2-r^2\sin^2\theta ~d\phi^2.\label{eq:Sch}
\ee
(We use geometric units wherein the speed of light in vacuum and the Newtonian constant of gravitation are set equal to unity.) Since the metric potentials in solution (\ref{eq:Sch}) do not depend on $t$ or $\phi$, the solution has 
\be
\underset{(1)}{A^\mu}=\frac{\partial x^\mu}{\partial t}, ~~~~ \underset{(2)}{A^\mu}=\frac{\partial x^\mu}{\partial \phi}~~~~~~ (x^\mu \equiv t,  r, \theta, \phi ~  \text{for}~\mu=0,1,2,3 ~\text{respectively}).\label{eq:Kil1}
\ee
as two Killing vectors. The spherical symmetry in (\ref{eq:Sch}) implies the existence of two additional Killing vectors
\be\label{eq:Kil2}
\left.\begin{aligned}
\underset{(3)}{A^\sigma}~\frac{\partial x^\mu}{\partial x^\sigma}=~\sin\phi \frac{\partial x^\mu}{\partial \theta} + \cot\theta\cos\phi~ \frac{\partial x^\mu}{\partial \phi},\\
\underset{(4)}{A^\sigma}~\frac{\partial x^\mu}{\partial x^\sigma}=-\cos\phi \frac{\partial x^\mu}{\partial \theta} + \cot\theta\sin\phi \frac{\partial x^\mu}{\partial \phi}.
 \end{aligned}
 \right\}
\ee
Since the linear combination of the Killing vectors is also a Killing vector, one can construct a `resultant' Killing field from these four fields by defining 
$A^\mu=a \underset{(1)}{A^\mu}+ b \underset{(2)}{A^\mu} + c \underset{(3)}{A^\mu} + d \underset{(4)}{A^\mu}$, where $a,b,c,d$ are constants. This gives
\be
A^\mu= (a,~ 0, ~ c \sin\phi - d \cos\phi, ~b+c \cot\theta\cos\phi+ d \cot\theta\sin\phi).
\ee
The definitions (\ref{eq:F}), (\ref{eq:U}) and (\ref{eq:VV}) then give the following non-vanishing independent components  of $F_{\mu\nu}$,  $U_{\mu\nu\alpha}$ and $V_{\mu\nu\alpha\kappa}$:

\be
\left.\begin{aligned}
& F_{tr} =\frac{a m}{r^2},~~~
F_{r\theta}= r(c \sin\phi - d \cos\phi),\\
& F_{r\phi}= r\sin\theta[\cos\theta(d \sin\phi + c \cos\phi)+b \sin\theta],\\
& F_{\theta\phi}=- r^2\sin\theta[\sin\theta(d\sin\phi +c\cos\phi)-b\cos\theta];
\end{aligned}
\right \}
\ee

\be
\left.\begin{aligned}
& U_{trr} =-\frac{2a m}{r^3},
~~~U_{t\theta t} =-\frac{m(1-2m/r)(c \sin\phi - d \cos\phi)}{r},
~~~U_{t\theta \theta} =\frac{a m(1-2m/r)}{r},\\
& U_{t\phi t} =-\frac{m(1-2m/r)\sin\theta[\cos\theta(d \sin\phi + c \cos\phi)+b \sin\theta]}{r},
~~~U_{t\phi \phi} =\frac{a m(1-2m/r)\sin^2\theta}{r},\\
& U_{r \theta r} =\frac{m(c \sin\phi - d \cos\phi)}{r(1-2m/r)},
~~~U_{r\phi r} =\frac{m\sin\theta[\cos\theta(d \sin\phi + c \cos\phi)+b \sin\theta]}{r(1-2m/r)},\\
& U_{\theta\phi \theta} =- m r[\sin2\theta(d \sin\phi + c \cos\phi)+2 b \sin^2\theta],
~~~U_{\theta\phi \phi} = 2m r(c \sin\phi - d  \cos\phi) \sin^2\theta];
\end{aligned}
\right \} \label{eq:U-Sch}
\ee

\be
\left.\begin{aligned}
& V_{trt\theta} =\frac{3 m(1-2m/r)(c \sin\phi - d \cos\phi)}{r^2},\\
& V_{trt\phi} =\frac{3 m(1-2m/r)\sin\theta[\cos\theta(d \sin\phi + c \cos\phi)+b \sin\theta]}{r^2},\\
& V_{r\theta\theta\phi} =- 3 m \sin\theta [\cos\theta(d \sin\phi + c \cos\phi)+b \sin\theta],\\
& V_{r\theta\theta\phi} = 3 m \sin^2\theta (c \sin\phi - d \cos\phi).
\end{aligned}
\right \} \label{eq:V-Sch}
\ee
It is already known that the non-vanishing components of the Weyl tensor for the spacetime (\ref{eq:Sch}) are given by
\be
\left.\begin{aligned}
& C_{trtr} =\frac{2 m}{r^3},~~~C_{t \theta t \theta} =-\frac{ m (1-2m/r)}{r},~~~C_{t \phi t \phi} =-\frac{ m (1-2m/r)\sin^2\theta}{r},\\
& C_{r \theta r \theta} =\frac{ m}{r (1-2m/r)},~~~C_{r \phi r \phi} =\frac{ m\sin^2\theta}{r (1-2m/r)},
~~~C_{\theta \phi \theta \phi} =-2 mr \sin^2\theta.
\end{aligned}
\right \} \label{eq:C-Sch}
\ee
It is now apparent that the tensor $V_{\mu\nu\kappa\alpha}$ depicted in (\ref{eq:V-Sch}) is distinct from the Weyl one described in  (\ref{eq:C-Sch}). Although the tensor $V_{\mu\nu\kappa\alpha}$ shares all the geometric symmetries of the Weyl tensor, as we have seen earlier, however its covariant divergence is not vanishing here, unlike the Weyl tensor, which has $C_{\mu\nu\kappa~~;\alpha}^{~~~~\alpha}=0$ for the Ricci-flat spacetimes. 
Consequently, the tensor $U_{\mu\nu\kappa}$, playing the role of the potential for $V_{\mu\nu\kappa\alpha}$ and depicted in (\ref{eq:U-Sch}), is distinct from the Lanczos tensor. The latter, for the spacetime (\ref{eq:Sch}), is given by \cite{Novello}
\be
\left.\begin{aligned}
& L_{trt} =\frac{2 m}{3 r^2},
~~~L_{r\theta \theta} =-\frac{ m}{3(1-2m/r)},
~~~L_{r\phi \phi} =-\frac{m\sin^2\theta}{3(1-2m/r)},\\
\end{aligned}
\right. \label{eq:L-Sch}
\ee
which do not match with  (\ref{eq:U-Sch}) for any possible values of the constants $a, b, c, d$. The values in (\ref{eq:L-Sch}) are calculated in the Lanczos gauge conditions $L_{\mu~~\nu}^{~~\nu}=0$, $L_{\mu\nu~;\sigma}^{~~~\sigma}=0$. Although $U_{\mu~~\nu}^{~~\nu}=0$ is admitted for the spacetime (\ref{eq:L-Sch}), as has been mentioned in  (\ref{eq:USym}), however the second condition is not satisfied, since
\be
\left.\begin{aligned}
& U^{~~\sigma}_{tr~;\sigma} =\frac{4 a m^2}{r^5},~~~U^{~~~\sigma}_{r\theta~~;\sigma} =-\frac{2 m (c \sin\phi - d \cos\phi)}{r^2},\\
& U^{~~~\sigma}_{r\phi~~;\sigma} =-\frac{2m\sin\theta[\cos\theta(d \sin\phi + c \cos\phi)+b \sin\theta]}{r^2},\\
& U^{~~~\sigma}_{\theta\phi~~;\sigma} =-\frac{4m\sin\theta[\sin\theta(d \sin\phi + c \cos\phi)- b \cos\theta]}{r}
\end{aligned}
\right \}
\ee
are non-vanishing for any possible non-trivial choice of the constants $a, b, c, d$.

\subsection{Example 2}

As another example, let us consider a non-Ricci-flat spacetime, for instance the one given by the Godel solution \cite{Novello}
\be
ds^2= dt^2 - dx^2 + \frac{1}{2} e^{2 a x} dy^2 - dz^2 + 2 e^{a x} dt ~dy, ~~~ a=\text{constant}.\label{eq:godel}
\ee
Since the metric potentials do not depend on $t$, $y$ and $z$, the spacetime has 
\be
\underset{(1)}{A^\mu}=\frac{\partial x^\mu}{\partial t}, ~~~ \underset{(2)}{A^\mu}=\frac{\partial x^\mu}{\partial y}, ~~~ \underset{(3)}{A^\mu}=\frac{\partial x^\mu}{\partial z}~~~~~~ (x^0 \equiv t, x^1\equiv x, x^2\equiv y, x^3\equiv z).
\ee
as Killing vector fields. Following the steps of the earlier example, we construct a `resultant' Killing field from the linear combination of these vector fields, which amounts to
\be
A^\mu= (\ell,~ 0, ~ p, ~n), ~~~~ \text{with}~ ~~\ell,~ p, ~n = \text{constant}.
\ee
The non-vanishing, independent components  of the tensors $F_{\mu\nu}$,  $U_{\mu\nu\alpha}$ and $V_{\mu\nu\alpha\kappa}$ then come out as

\be
 F_{tx} =\frac{1}{2}pa~e^{ax},~~~
 F_{xy} =-\frac{1}{2}a~e^{ax}\left(\ell+p~e^{ax}\right);
\ee

\be
\left.\begin{aligned}
& U_{txx} =\frac{1}{2}a^2\left(\ell+p~e^{ax}\right),~~~
 U_{tyt} =-\frac{1}{4}p~a^2 e^{2ax},\\
& U_{tyy} =\frac{1}{4}\ell~a^2 e^{2ax},~~~
 U_{xyx} =-\frac{1}{4}a^2 e^{ax} \left(2\ell+3p~e^{ax}\right);\\
\end{aligned}
\right \} \label{eq:Ugodel}
\ee

\be
V_{tyxy}=\frac{1}{8}p~a^3 e^{3ax},~~~ V_{tzxz}=-\frac{1}{4}p~a^3 e^{ax},~~~V_{xzyz}=-\frac{1}{4}p~a^3 e^{2ax}.
\ee
The corresponding non-vanishing, independent components  of the tensor $L_{\mu\nu\alpha}$, for the spacetime (\ref{eq:godel}), have been calculated in \cite{Novello} as
\be
L_{txy}=\frac{a}{18}e^{ax}=-L_{tyx}=-\frac{L_{xyt}}{2}=-\frac{L_{xyy}}{3}; \label{eq:L-godel}
\ee
While the components of the tensor $L_{\mu\nu\alpha}$ in (\ref{eq:L-godel}) admit the Lanczos gauge conditions, those in (\ref{eq:Ugodel}) do not admit the corresponding conditions for $U_{\mu\nu\alpha}$. (i.e., $U_{\mu~~\nu}^{~~\nu}\neq 0$, $U_{\mu\nu~;\sigma}^{~~~\sigma}\neq 0$ for (\ref{eq:Ugodel})).
Similarly, one can calculate the Weyl tensor $C_{\mu\nu\alpha\kappa}$ for the spacetime (\ref{eq:godel}), which comes out as
\be
\left.\begin{aligned}
& C_{txtx} =-\frac{a^2}{6}=-\frac{C_{tztz}}{2}=- C_{xzxz},\\
& C_{txxy} =\frac{a^2}{6} e^{ax}=\frac{ C_{tzyz}}{2},\\
& C_{tyty} =-\frac{a^2}{12} e^{2ax}=\frac{ C_{xyxy}}{4}=-\frac{ C_{yzyz}}{5}.
\end{aligned}
\right \} 
\ee 
Clearly the tensors $U_{\mu\nu\alpha}$ and $V_{\mu\nu\alpha\kappa}$ do not match with $L_{\mu\nu\alpha}$ and $C_{\mu\nu\alpha\kappa}$.

\clearpage

\appendix
\section{Double two-form and its duals}

For any double two-form tensor $D_{\mu\nu\kappa\alpha}$, which by definition satisfies
\be
D_{\mu\nu\kappa\alpha}=-D_{\nu\mu\kappa\alpha}=-D_{\mu\nu\alpha\kappa},  \label{eq:S1}
\ee
its left-dual and right-dual are defined respectively by \cite{MTW}:
\be
{*D}_{\mu\nu\kappa\alpha}=\frac{1}{2}e_{\mu\nu\sigma\tau}D^{\sigma\tau}_{~~~\kappa\alpha}, ~~~~~~~
{D*}_{\mu\nu\kappa\alpha}=\frac{1}{2}e_{\kappa\alpha\sigma\tau}D^{~~~\sigma\tau}_{\mu\nu}.   \label{eq:d}
\ee
Here $e_{\mu\nu\kappa\alpha}$ is the totally antisymmetric Levi-Civita tensor defined by 
\[
e_{\mu\nu\kappa\alpha}=\sqrt{-g}~\epsilon_{\mu\nu\kappa\alpha}, ~~~ e^{\mu\nu\kappa\alpha}=\frac{-1}{\sqrt{-g}}~\epsilon^{\mu\nu\kappa\alpha},
~~~~ g=\text{det} (g_{\mu\nu}),
\]
where $\epsilon_{\mu\nu\kappa\alpha}$ is given by
\[
\epsilon_{\mu\nu\kappa\alpha}=\left\{
\begin{aligned}
 0 & ~~~~~\text{if any two indices are equal}; \\
  1  &~~~ ~~\text{if}~ (\mu\nu\kappa\alpha)~  \text{is an even permutation of} ~(1, 2, 3, 0);\\
   -1&   ~~ ~~ ~\text{if}~ (\mu\nu\kappa\alpha) ~ \text{is an odd permutation of}~ (1, 2, 3, 0);\\
\end{aligned} 
\right.
%\label{eq:U-godel}
\]
and  $\epsilon^{\mu\nu\kappa\alpha}$ has the same numerical values as $\epsilon_{\mu\nu\kappa\alpha}$ has.
From these definitions, one may check that
\be
- e_{\mu_1\mu_2\mu_3\mu_4} ~ e^{\lambda_1\lambda_2\lambda_3\lambda_4}= 
  \left| {\begin{array}{cccc}
 \delta^{\lambda_1}_{\mu_1}  &  \delta^{\lambda_1}_{\mu_2}  & \delta^{\lambda_1}_{\mu_3}  & \delta^{\lambda_1}_{\mu_4} \\
 \delta^{\lambda_2}_{\mu_1}  &  \delta^{\lambda_2}_{\mu_2}  & \delta^{\lambda_2}_{\mu_3}  & \delta^{\lambda_2}_{\mu_4} \\
 \delta^{\lambda_3}_{\mu_1}  &  \delta^{\lambda_3}_{\mu_2}  & \delta^{\lambda_3}_{\mu_3}  & \delta^{\lambda_3}_{\mu_4} \\
 \delta^{\lambda_4}_{\mu_1}  &  \delta^{\lambda_4}_{\mu_2}  & \delta^{\lambda_4}_{\mu_3}  & \delta^{\lambda_4}_{\mu_4} \\
  \end{array} } \right|
\equiv \delta^{\lambda_1\lambda_2\lambda_3\lambda_4}_{\mu_1\mu_2\mu_3\mu_4},\label{eq:LC1}
\ee
\be       
-e_{\mu_1 . .\mu_n \mu_{n+1}..\mu_4} ~ e^{\mu_1 . .\mu_n\lambda_{n+1}..\lambda_4}=\delta_{\mu_1 . .\mu_n \mu_{n+1}..\mu_4}^{\mu_1 . .\mu_n\lambda_{n+1}..\lambda_4}= n!~\delta_{\mu_{n+1}..\mu_4}^{\lambda_{n+1}..\lambda_4}
=n!
 \left| {\begin{array}{cccc}
 \delta^{\lambda_{n+1}}_{\mu_{n+1}}~ . & .  & \delta^{\lambda_{n+1}}_{\mu_4} \\
.& . . & .  & \\
.& . . & .  & \\
 \delta^{\lambda_4}_{\mu_{n+1}} ~. & .  & \delta^{\lambda_4}_{\mu_4} \\
 \end{array} } \right|,
 \label{eq:LC2}
\ee
where $n\leq 4$.
From the definitions (\ref{eq:S1}) and (\ref{eq:d}), it is apparent that 
\be
{*D}_{\mu\nu\kappa\alpha}={D*}_{\kappa\alpha\mu\nu},   \label{eq:LRd}
\ee
if the tensor is a symmetric double two-form, i.e., it also satisfies $D_{\mu\nu\kappa\alpha}=D_{\kappa\alpha\mu\nu}$, in addition to the symmetries (\ref{eq:S1}). 
Similarly, the double-dual of the tensor is defined by
\be
{*D*}_{\mu\nu}^{~~~\kappa\alpha}=\frac{1}{4} e_{\mu\nu\sigma\tau} ~ e^{\kappa\alpha\rho\lambda} D^{\sigma\tau}_{~~~\rho\lambda}=- \delta^{\kappa\alpha\rho\lambda}_{\mu\nu\sigma\tau} ~D^{\sigma\tau}_{~~~\rho\lambda},   \label{eq:dd}
\ee
by virtue of (\ref{eq:LC1}).
One can check that $**=-1$. By using (\ref{eq:LC1}) in (\ref{eq:dd}), one can derive the Lanczos identity for a symmetric double two-form, which reads
\begin{eqnarray}\nonumber
{*D*}_{\mu\nu\kappa\alpha} &=& - D_{\mu\nu\kappa\alpha}+ g_{\nu\alpha} \left(D_{\mu\kappa}-\frac{1}{4}D g_{\mu\kappa}\right)+ g_{\mu\kappa} \left(D_{\nu\alpha}-\frac{1}{4}D g_{\nu\alpha}\right)\\
&&- g_{\nu\kappa} \left(D_{\mu\alpha}-\frac{1}{4}D g_{\mu\alpha}\right)- g_{\mu\alpha} \left(D_{\nu\kappa}-\frac{1}{4}D g_{\nu\kappa}\right),
\end{eqnarray}
where $D_{\mu\nu}\equiv D^\sigma_{~\mu\sigma\nu}$ and $D \equiv D^\sigma_{~\sigma}$.

\section{Symmetries of a symmetric double two-form and its dual}

As $V_{\mu\nu\kappa\alpha}$ is a symmetric double two-form (by virtue of (\ref{eq:I-V})), its duals can be defined, following (\ref{eq:d}), which implies
\be
{*V}_{\mu\nu\kappa\alpha}=-{*V}_{\nu\mu\kappa\alpha}=-{*V}_{\mu\nu\alpha\kappa}.  \label{eq:Vd1}
\ee
In view of (\ref{eq:LRd}), we also have
\be
{*V}_{\mu\nu\kappa\alpha}={V*}_{\kappa\alpha\mu\nu}.   \label{eq:LRV}
\ee
Since $**=-1$, we can write
\[
-V^{\mu\sigma}_{~~~\rho\alpha}={**V}^{\mu\sigma}_{~~~\rho\alpha}=\frac{1}{2}e^{\mu\sigma\nu\kappa}{*V}_{\nu\kappa\rho\alpha}.
\]
From this, we can write the trace of the tensor as
\[
-V^{\mu\sigma}_{~~~\mu\alpha}= \frac{1}{2}e^{\mu\sigma\nu\kappa}{*V}_{\nu\kappa\mu\alpha}.
\]
As the dummy indices $\mu,\nu,\kappa$ appearing in this identity can be permuted arbitrarily, we write out all the three even permutation and add the resulting equations, giving
\[
-V^{\mu\sigma}_{~~~\mu\alpha}= \frac{1}{6}\left(e^{\mu\sigma\nu\kappa}{*V}_{\nu\kappa\mu\alpha}+e^{\nu\sigma\kappa\mu}{*V}_{\kappa\mu\nu\alpha}+e^{\kappa\sigma\mu\nu}{*V}_{\mu\nu\kappa\alpha}\right).
\]
By using the symmetry properties of $e^{\mu\nu\kappa\sigma}$, this can be written as
\[
-V^{\mu\sigma}_{~~~\mu\alpha}= \frac{1}{6}e^{\mu\sigma\nu\kappa}({*V}_{\nu\kappa\mu\alpha}+{*V}_{\kappa\mu\nu\alpha}+{*V}_{\mu\nu\kappa\alpha}),
\]
implying
\be
V^\nu_{~\mu\nu\alpha}=0 ~\Leftrightarrow ~{*V}_{\mu\nu\kappa\alpha}+{*V}_{\nu\kappa\mu\alpha}+{*V}_{\kappa\mu\nu\alpha}=0.  \label{eq:VdCyclic}
\ee
By recalling that the tensor $V_{\mu\nu\kappa\alpha}$ is indeed trace-free, one then obtains a cyclic symmetry for its dual:
\be
{*V}_{\mu\nu\kappa\alpha}+{*V}_{\nu\kappa\mu\alpha}+{*V}_{\kappa\mu\nu\alpha}=0.  \label{eq:Vd2}
\ee
Like the tensor $V_{\mu\nu\kappa\alpha}$, its dual too constitutes a symmetric double two-form.  This can be shown as the following. By virtue of (\ref{eq:Vd1}),  the identity (\ref{eq:Vd2}) yields
\[
{*V}_{\mu\nu\kappa\alpha}=- {*V}_{\nu\kappa\mu\alpha} - {*V}_{\kappa\mu\nu\alpha}={*V}_{\nu\kappa\alpha\mu} + {*V}_{\kappa\mu\alpha\nu}.
\]
By the use of  (\ref{eq:Vd2}) again, this yields
\[
{*V}_{\mu\nu\kappa\alpha}=-{*V}_{\kappa\alpha\nu\mu}- {*V}_{\alpha\nu\kappa\mu}- {*V}_{\mu\alpha\kappa\nu}- {*V}_{\alpha\kappa\mu\nu}
=2 {*V}_{\kappa\alpha\mu\nu}+ {*V}_{\alpha\nu\mu\kappa}+{*V}_{\mu\alpha\nu\kappa},
\]
which reduces, by virtue of (\ref{eq:Vd2}), to
\[
{*V}_{\mu\nu\kappa\alpha}
=2 {*V}_{\kappa\alpha\mu\nu}- {*V}_{\nu\mu\alpha\kappa},
\]
 implying
\be
{*V}_{\mu\nu\kappa\alpha} = {*V}_{\kappa\alpha\mu\nu}.  \label{eq:Vd3}
\ee
This, taken together with the symmetries encoded in (\ref{eq:Vd1}), garnishes ${*V}_{\mu\nu\kappa\alpha}$ with the status of a symmetric double two-form.  One also notes that by virtue of  (\ref{eq:LRV}), the identity (\ref{eq:Vd3}) amounts to
\[
{*V}_{\kappa\alpha\mu\nu} = {V*}_{\kappa\alpha\mu\nu},  \label{eq:Vd4}
\]
which is consistent with our earlier finding (\ref{eq:LRV1}) derived from the definition (\ref{eq:V}) of the tensor $V_{\mu\nu\kappa\alpha}$.
In a way similar to (\ref{eq:VdCyclic}), by starting with the trace of ${*V}_{\mu\nu\kappa\alpha}$, one can show that
\be
{*V}^\nu_{~\mu\nu\alpha}=0 ~\Leftrightarrow ~V_{\mu\nu\kappa\alpha}+V_{\nu\kappa\mu\alpha}+V_{\kappa\mu\nu\alpha}=0.  \label{eq:VCyclic}
\ee
In view of the cyclic  symmetry for $V_{\mu\nu\kappa\alpha}$ admitted in (\ref{eq:CI-V}), the trace-freedom for its dual is thus ascertained.

\section{Some identities for a symmetric double two-form}

By using the definition (\ref{eq:d}), one can write 
\[
{*V}_{\mu~~~\kappa}^{~~\sigma\rho} ~ {*V}_{\nu\sigma\rho\alpha} = g_{\mu\tau} ~ g_{\kappa\gamma}~ {*V}^{\tau\sigma\rho\gamma}~ {*V}_{\nu\sigma\rho\alpha} = g_{\mu\tau} ~ g_{\kappa\gamma}~ \frac{1}{2}e^{\tau\sigma\beta\theta} ~V_{\beta\theta}^{~~\rho\gamma}~\frac{1}{2} e_{\nu\sigma\eta\lambda}~ V_{~~~\rho\alpha}^{\eta\lambda}. 
\]
By the use of (\ref{eq:LC2}), i.e., $e^{\tau\sigma\beta\theta}~ e_{\nu\sigma\eta\lambda}=-\delta^{\tau\beta\theta}_{\nu\eta\lambda}=-
  \left| {\begin{array}{ccc}
 \delta^\tau_\nu  &    \delta^\tau_\eta  &    \delta^\tau_\lambda   \\
 \delta^\beta_\nu  &    \delta^\beta_\eta  &    \delta^\beta_\lambda   \\
 \delta^\theta_\nu\ &    \delta^\theta_\eta  &    \delta^\theta_\lambda   \\
  \end{array} } \right|$, 
the above equation reduces to
\be
{*V}_{\mu~~~\kappa}^{~~\sigma\rho} ~ {*V}_{\nu\sigma\rho\alpha} =V_{\mu~~~\alpha}^{~\sigma\rho} ~ V_{\nu\sigma\rho\kappa}  - \frac{1}{2}g_{\mu\nu}~V_{\kappa\sigma\rho\tau}V_{\alpha}^{~\sigma\rho\tau}.  \label{eq:C1}
\ee

\bigskip
\noindent
By using the identities (\ref{eq:CI-V}) and (\ref{eq:I-V}), one can write
\begin{eqnarray}\nonumber
V_{\alpha\beta}^{~~~\sigma\rho}~V_{\mu\nu\sigma\rho}
&=& \left(- V_{\beta~\alpha}^{~\sigma~\rho} - V_{~\alpha\beta}^{\sigma~~\rho}\right) \left(-V_{\nu\sigma\mu\rho}-V_{\sigma\mu\nu\rho}\right)
 = \left( V_{\beta~~~\alpha}^{~\sigma\rho} - V_{\alpha~~~\beta}^{~\sigma\rho}\right) \left(V_{\mu\rho\sigma\nu} - V_{\mu\sigma\rho\nu}\right)\\ \nonumber
&&= V_{\beta~~~\alpha}^{~\sigma\rho} ~ V_{\mu\rho\sigma\nu} +  V_{\alpha~~~\beta}^{~\sigma\rho}~V_{\mu\sigma\rho\nu}
- V_{\beta~~~\alpha}^{~\sigma\rho} ~V_{\mu\sigma\rho\nu} -  V_{\alpha~~~\beta}^{~\sigma\rho}~V_{\mu\rho\sigma\nu} \\ \nonumber
&&= V_{\alpha~~\beta}^{~\rho\sigma} ~ V_{\mu\rho\sigma\nu} +  V_{\alpha~~~\beta}^{~\sigma\rho}~V_{\mu\sigma\rho\nu}
- V_{\beta~~~\alpha}^{~\sigma\rho} ~V_{\mu\sigma\rho\nu} -  V_{\beta~~\alpha}^{~\rho\sigma}~V_{\mu\rho\sigma\nu}\\
&&= 2\left(V_{\alpha~~\beta}^{~\rho\sigma} ~ V_{\mu\rho\sigma\nu} -  V_{\beta~~\alpha}^{~\rho\sigma}~V_{\mu\rho\sigma\nu}\right),  \label{eq:C2}
\end{eqnarray}
which is obtained by renaming some dummy indices appropriately in the last step. As ${*V}_{\mu\nu\kappa\alpha}$ possesses all the symmetries of $V_{\mu\nu\kappa\alpha}$,  one can similarly write
\be
{*V}_{\alpha\beta}^{~~~~\sigma\rho}~{*V}_{\mu\nu\sigma\rho}= 2\left({*V}_{\alpha~~~\beta}^{~~\rho\sigma} ~ {*V}_{\mu\rho\sigma\nu} -  {*V}_{\beta~~~\alpha}^{~~\rho\sigma}~{*V}_{\mu\rho\sigma\nu}\right).  \label{eq:C3}
\ee
By adding  (\ref{eq:C2}) and (\ref{eq:C3}), we then have
\[
V_{\alpha\beta}^{~~~\sigma\rho}~V_{\mu\nu\sigma\rho} + {*V}_{\alpha\beta}^{~~~~\sigma\rho}~{*V}_{\mu\nu\sigma\rho}
= 2\left[V_{\alpha~~\beta}^{~\rho\sigma} ~ V_{\mu\rho\sigma\nu} + {*V}_{\alpha~~\beta}^{~\rho\sigma} ~ {*V}_{\mu\rho\sigma\nu} 
- \left( V_{\beta~~\alpha}^{~\rho\sigma}~V_{\mu\rho\sigma\nu} + {*V}_{\beta~~\alpha}^{~\rho\sigma}~{*V}_{\mu\rho\sigma\nu}\right)\right], 
\]
which can be written as
\be
V_{\alpha\beta}^{~~~\sigma\rho}~V_{\mu\nu\sigma\rho} + {*V}_{\alpha\beta}^{~~~~\sigma\rho}~{*V}_{\mu\nu\sigma\rho}
= 2\left[W_{\alpha\mu\beta\nu} - W_{\beta\mu\alpha\nu} \right],  \label{eq:C4}
\ee
by virtue of (\ref{eq:W}).  By using (\ref{eq:LRV1}), one can write
\begin{eqnarray}\nonumber
{*V}_{\alpha\beta}^{~~~~\sigma\rho}~{*V}_{\mu\nu\sigma\rho}
&=& {V*}_{\alpha\beta}^{~~~~\sigma\rho}~{V*}_{\mu\nu\sigma\rho}
=\frac{1}{4}e^{\sigma\rho\tau\lambda}e_{\sigma\rho\kappa\gamma}~V_{\alpha\beta\tau\lambda}~V_{\mu\nu}^{~~~\kappa\gamma}
=-\frac{1}{2}\delta^{\tau\lambda}_{\kappa\gamma}~V_{\alpha\beta\tau\lambda}~V_{\mu\nu}^{~~~\kappa\gamma}\\
&& =-\frac{1}{2}\left(\delta^\tau_\kappa\delta^\lambda_\gamma - \delta^\tau_\gamma\delta^\lambda_\kappa \right)V_{\alpha\beta\tau\lambda}~V_{\mu\nu}^{~~~\kappa\gamma}
=- V_{\alpha\beta}^{~~~\kappa\gamma}~V_{\mu\nu\kappa\gamma},   \label{eq:C5}
\end{eqnarray}
by the use of (\ref{eq:LC2}). By virtue of  (\ref{eq:C5}), the left hand side of  (\ref{eq:C4}) then vanishes identically, implying
\be
W_{\alpha\mu\beta\nu} = W_{\beta\mu\alpha\nu}.  \label{eq:C6}
\ee

\end{document}